\documentclass[12pt,superscript,nomove]{article} 
\usepackage{cite}
\usepackage{graphics,graphicx}

\newcommand{\degree}{$^{\circ}$}

\newcommand{\Fermi}{{\it Fermi}}
\newcommand{\RXTE}{{\it RXTE}}
\newcommand{\Swift}{{\it Swift}}

\begin{document}
\pagestyle{plain}
\begin{center}
{\Large \bf A change in the optical polarization associated 
with a $\gamma$-ray flare in the blazar 3C 279}
\end{center}

\noindent
A.~A.~Abdo$^{1,2}$, 
M.~Ackermann$^{3}$, 
M.~Ajello$^{3}$, 
M.~Axelsson$^{4,5}$, 
L.~Baldini$^{6}$, 
J.~Ballet$^{7}$, 
G.~Barbiellini$^{8,9}$, 
D.~Bastieri$^{10,11}$, 
B.~M.~Baughman$^{12}$, 
K.~Bechtol$^{3}$
% ...

\medskip
\begin{enumerate}
\item[1.] Space Science Division, Naval Research Laboratory, Washington, DC 20375, USA
\item[2.] National Research Council Research Associate, National Academy of Sciences, Washington, DC 20001, USA
\item[3.] W. W. Hansen Experimental Physics Laboratory, Kavli Institute for Particle Astrophysics and Cosmology, Department of Physics and SLAC National Accelerator Laboratory, Stanford University, Stanford, CA 94305, USA
\item[4.] Department of Astronomy, Stockholm University, SE-106 91 Stockholm, Sweden
\item[5.] The Oskar Klein Centre for Cosmoparticle Physics, AlbaNova, SE-106 91 Stockholm, Sweden
\item[6.] Istituto Nazionale di Fisica Nucleare, Sezione di Pisa, I-56127 Pisa, Italy
\item[7.] Laboratoire AIM, CEA-IRFU/CNRS/Universit\'e Paris Diderot, Service d'Astrophysique, CEA Saclay, 91191 Gif sur Yvette, France
\item[8.] Istituto Nazionale di Fisica Nucleare, Sezione di Trieste, I-34127 Trieste, Italy
\item[9.] Dipartimento di Fisica, Universit\`a di Trieste, I-34127 Trieste, Italy
\item[10.] Istituto Nazionale di Fisica Nucleare, Sezione di Padova, I-35131 Padova, Italy
\item[11.] Dipartimento di Fisica ``G. Galilei", Universit\`a di Padova, I-35131 Padova, Italy
\item[12.] Department of Physics, Center for Cosmology and Astro-Particle Physics, The Ohio State University, Columbus, OH 43210, USA
\item[13.] Combined Array for Research in Millimeter-wave Astronomy (CARMA), Big Pine, CA 93514, USA
\item[14.] Radio Astronomy Laboratory, University of California, Berkeley, CA 94720, USA
\item[15.] Istituto Nazionale di Fisica Nucleare, Sezione di Perugia, I-06123 Perugia, Italy
\item[16.] Dipartimento di Fisica, Universit\`a degli Studi di Perugia, I-06123 Perugia, Italy
\item[17.] Dipartimento di Fisica ``M. Merlin" dell'Universit\`a e del Politecnico di Bari, I-70126 Bari, Italy
\item[18.] Istituto Nazionale di Fisica Nucleare, Sezione di Bari, 70126 Bari, Italy
\item[19.] Laboratoire Leprince-Ringuet, \'Ecole polytechnique, CNRS/IN2P3, Palaiseau, France
\item[20.] Department of Physics, University of Washington, Seattle, WA 98195-1560, USA
\item[21.] Institut de Ciencies de l'Espai (IEEC-CSIC), Campus UAB, 08193 Barcelona, Spain
\item[22.] INAF-Istituto di Astrofisica Spaziale e Fisica Cosmica, I-20133 Milano, Italy
\item[23.] Agenzia Spaziale Italiana (ASI) Science Data Center, I-00044 Frascati (Roma), Italy
\item[24.] NASA Goddard Space Flight Center, Greenbelt, MD 20771, USA
\item[25.] Center for Research and Exploration in Space Science and Technology (CRESST) and NASA Goddard Space Flight Center, Greenbelt, MD 20771, USA
\item[26.] Department of Physics and Center for Space Sciences and Technology, University of Maryland Baltimore County, Baltimore, MD 21250, USA
\item[27.] George Mason University, Fairfax, VA 22030, USA
\item[28.] Laboratoire de Physique Th\'eorique et Astroparticules, Universit\'e Montpellier 2, CNRS/IN2P3, Montpellier, France
\item[29.] Max-Planck Institut f\"ur extraterrestrische Physik, 85748 Garching, Germany
\item[30.] Department of Physics and Astronomy, Sonoma State University, Rohnert Park, CA 94928-3609, USA
\item[31.] Department of Physics, Stockholm University, AlbaNova, SE-106 91 Stockholm, Sweden
\item[32.] Royal Swedish Academy of Sciences Research Fellow, funded by a grant from the K. A. Wallenberg Foundation
\item[33.] Institut universitaire de France, 75005 Paris, France
\item[34.] Dipartimento di Fisica, Universit\`a di Udine and Istituto Nazionale di Fisica Nucleare, Sezione di Trieste, Gruppo Collegato di Udine, I-33100 Udine, Italy
\item[35.] CNRS/IN2P3, Centre d'\'Etudes Nucl\'eaires Bordeaux Gradignan, UMR 5797, Gradignan, 33175, France
\item[36.] Universit\'e de Bordeaux, Centre d'\'Etudes Nucl\'eaires Bordeaux Gradignan, UMR 5797, Gradignan, 33175, France
\item[37.] Osservatorio Astronomico di Trieste, Istituto Nazionale di Astrofisica, I-34143 Trieste, Italy
\item[38.] Max-Planck-Institut f\"ur Radioastronomie, Auf dem H\"ugel 69, 53121 Bonn, Germany
\item[39.] Department of Physical Sciences, Hiroshima University, Higashi-Hiroshima, Hiroshima 739-8526, Japan
\item[40.] Department of Astronomy and Astrophysics, Pennsylvania State University, University Park, PA 16802, USA
\item[41.] Department of Physics and Department of Astronomy, University of Maryland, College Park, MD 20742, USA
\item[42.] INAF Istituto di Radioastronomia, 40129 Bologna, Italy
\item[43.] Center for Space Plasma and Aeronomic Research (CSPAR), University of Alabama in Huntsville, Huntsville, AL 35899, USA
\item[44.] Department of Physics, Royal Institute of Technology (KTH), AlbaNova, SE-106 91 Stockholm, Sweden
\item[45.] Dr. Remeis-Sternwarte Bamberg, Sternwartstrasse 7, D-96049 Bamberg, Germany
\item[46.] Erlangen Centre for Astroparticle Physics, D-91058 Erlangen, Germany
\item[47.] Universities Space Research Association (USRA), Columbia, MD 21044, USA
\item[48.] Research Institute for Science and Engineering, Waseda University, 3-4-1, Okubo, Shinjuku, Tokyo, 169-8555 Japan
\item[49.] Department of Physics, Tokyo Institute of Technology, Meguro City, Tokyo 152-8551, Japan
\item[50.] Cosmic Radiation Laboratory, Institute of Physical and Chemical Research (RIKEN), Wako, Saitama 351-0198, Japan
\item[51.] Centre d'\'Etude Spatiale des Rayonnements, CNRS/UPS, BP 44346, F-30128 Toulouse Cedex 4, France
\item[52.] ICRAR/Curtin Institute of Radio Astronomy, Bentley WA 6102., Australia
\item[53.] Cahill Center for Astronomy and Astrophysics, California Institute of Technology, Pasadena, CA 91125, USA
\item[54.] Istituto Nazionale di Fisica Nucleare, Sezione di Roma ``Tor Vergata", I-00133 Roma, Italy
\item[55.] Department of Physics and Astronomy, University of Denver, Denver, CO 80208, USA
\item[56.] Department of Physics, Graduate School of Science, University of Tokyo, 7-3-1 Hongo, Bunkyo-ku, Tokyo 113-0033, Japan
\item[57.] Santa Cruz Institute for Particle Physics, Department of Physics and Department of Astronomy and Astrophysics, University of California at Santa Cruz, Santa Cruz, CA 95064, USA
\item[58.] Institut f\"ur Astro- und Teilchenphysik and Institut f\"ur Theoretische Physik, Leopold-Franzens-Universit\"at Innsbruck, A-6020 Innsbruck, Austria
\item[59.] Kavli Institute for Cosmological Physics, University of Chicago, Chicago, IL 60637, USA
\item[60.] Space Sciences Division, NASA Ames Research Center, Moffett Field, CA 94035-1000, USA
\item[61.] NYCB Real-Time Computing Inc., Lattingtown, NY 11560-1025, USA
\item[62.] Department of Chemistry and Physics, Purdue University Calumet, Hammond, IN 46323-2094, USA
\item[63.] Institute of Space and Astronautical Science, JAXA, 3-1-1 Yoshinodai, Sagamihara, Kanagawa 229-8510, Japan
\item[64.] University of New Mexico, MSC07 4220, Albuquerque, NM 87131, USA
\item[65.] Partially supported by the International Doctorate on Astroparticle Physics (IDAPP) program
\item[66.] Instituci\'o Catalana de Recerca i Estudis Avan\c{c}ats (ICREA), Barcelona, Spain
\item[67.] Consorzio Interuniversitario per la Fisica Spaziale (CIFS), I-10133 Torino, Italy
\item[68.] Dipartimento di Fisica, Universit\`a di Roma ``Tor Vergata", I-00133 Roma, Italy
\item[69.] Space Science Institute, Boulder, CO 80301, USA
\item[70.] School of Pure and Applied Natural Sciences, University of Kalmar, SE-391 82 Kalmar, Sweden
\item[71.] Instituto de Astrofisica de Andaluc\'ia, CSIC, 18080 Granada, Spain
\item[72.] Department of Astronomy, University of Michigan, Ann Arbor, MI 48109-1042, USA
\item[73.] Pulkovo Observatory, 196140 St. Petersburg, Russia
\item[74.] Instituto de Astronom\'ia, Universidad Nacional Aut\'onoma de M\'exico, CP 04510 M\'exico, D. F., M\'exico
\item[75.] Tuorla Observatory, University of Turku, FI-21500 Piikki\"o, Finland
\item[76.] Department of Physics and Astronomy, Ohio University, Athens, OH 45701, USA
\item[77.] Osservatorio Astrofisico di Catania, 95123 Catania, Italy
\item[78.] Graduate Institute of Astronomy, National Central University, Jhongli 32054, Taiwan
\item[79.] INAF-Osservatorio Astronomico di Collurania, 64100 Teramo, Italy
\item[80.] Astronomical Institute, St. Petersburg State University, St. Petersburg, Russia
\item[81.] Cavendish Laboratory, Cambridge CB3 0HE, UK
\item[82.] Harvard-Smithsonian Center for Astrophysics, Cambridge, MA 02138, USA
\item[83.] Landessternwarte, Universit\"at Heidelberg, K\"onigstuhl, D 69117 Heidelberg, Germany
\item[84.] Instituto de Astronom\'ia, Universidad Nacional Aut\'onoma de M\'exico, CP 22860 Ensenada, B. C., M\'exico
\item[85.] Aalto University Mets\"ahovi Radio Observatory, FIN--02540 Kylm\"al\"a, Finland
\item[86.] Institute for Astrophysical Research, Boston University, Boston, MA 02215, USA
\item[87.] Hiroshima Astrophysical Science Center, Hiroshima University, Higashi-Hiroshima, Hiroshima 739-8526, Japan
\item[88.] Department of Physics and Astrophysics, Nagoya University, Chikusa-ku Nagoya 464-8602, Japan
\item[89.] Abastumani Observatory, Mt. Kanobili, 0301 Abastumani, Georgia
\item[90.] School of Physics and Astronomy, University of Southampton, Southampton SO17 BJ, UK
\item[91.] Nicolaus Copernicus Astronomical Center, 00-716 Warsaw, Poland
\item[92.] Osservatorio Astronomico di Roma, 00040 Monte Porzio Catone, Italy
\item[93.] INAF, Osservatorio Astronomico di Torino, I-10025 Pino Torinese (TO), Italy
\item[94.] Astrophyisikalisches Institut Potsdam, Potsdam, Germany
\item[95.] Astronomical Observatory, Jagiellonian University, 30-244 Krak\'ow, Poland
\end{enumerate}

{\bf 

It is widely accepted that strong and variable radiation
detected over all accessible energy bands
in a number of active galaxies arises from a
relativistic, Doppler-boosted jet pointing close to our line of
sight~\cite{MH97}.  
The size of the emitting zone and the location of this region 
relative to the central supermassive black hole are, however, 
poorly known, with estimates ranging from light-hours to a light-year or more. 
Here we report the coincidence of a gamma ($\gamma$)-ray flare with 
a dramatic change of optical polarization angle. 
This provides evidence for co-spatiality of optical and $\gamma$-ray 
emission regions and indicates a highly ordered jet magnetic field. 
The results also require a non-axisymmetric structure of the emission zone, 
implying a curved trajectory for the emitting material within the jet, 
with the dissipation region located at a considerable distance 
from the black hole, at about $10^{5}$ gravitational radii.

}

\bigskip

The flat spectrum radio quasar 3C~279 was the 
first bright $\gamma$-ray blazar reported by the EGRET 
instrument aboard the Compton Gamma-Ray Observatory 
to show strong and rapidly variable $\gamma$-ray 
emission~\cite{Har92, Kni93, Weh98};  recently, it also has been 
detected at photon energies above 100 GeV by the MAGIC ground-based 
Cherenkov telescope~\cite{MAGIC}.  
This blazar, at the redshift $z = 0.536$, harbors 
a black hole with mass~\cite{Woo02,Nilsson09} $M \simeq (3 - 8) \times 10^8 M_{\odot}$ 
(where $M_{\odot}$ is the mass of the Sun);  for specificity, we adopt 
$6 \times 10^8 M_{\odot}$.  It shows superluminal 
expansion best described as the jet material propagating with the 
bulk Lorentz factor $\Gamma_{\rm jet}  = 16 \pm 3$ at a 
small angle ($\theta \sim 2^{\circ}$) to our line of sight~\cite{Jor05}.  
The high degree of the optical polarization provides evidence for the presence 
of a well ordered magnetic field in the emission zone\cite{Lar08}.  This may  
reflect either the global topology of the large-scale magnetic field, or may result 
from the compression of chaotic magnetic fields in shocks and shear 
regions along the outflow~\cite{Lai80}. 

\bigskip

The best coverage of the broad-band flux variability of 
3C~279 has been obtained after the start of routine scientific operation 
of the Large Area Telescope (LAT)~\cite{LAT} onboard of the 
recently launched {\it Fermi} Gamma-ray Space Telescope 
(August 4th 2008 = 54682 Modified Julian Day, or MJD).  
In Figure~1 we plot the flux history in the 
$\gamma$-ray band above 200~MeV, as well as in the X-ray, optical, infrared, and radio 
bands together with polarization information in the optical band.  
Among all the observed bands, the $\gamma$-ray band shows the most violent 
variations, with a change by an order of magnitude in flux during the observation.  It also 
dominates the electromagnetic output of 3C~279, with the apparent $\gamma$-ray 
luminosity as much as $\sim 10^{48}$ erg s$^{-1}$ 
(see Figure~2 and ref.~\citen{Kni93, Weh98}).  
After being in the quiescent state for the first 100 days or so, the $\gamma$-ray flux 
starts to increase at $\sim$ 54780~MJD, but without any 
significant spectral changes:  the $\gamma$-ray 
photon index is relatively constant during the entire observed period.  
The high $\gamma$-flux state persists for $\sim$ 120 days and is associated 
with erratic flaring, accompanied by bright and variable optical emission. 

\bigskip

Towards the end of the high-flux state there is a sharp $\gamma$-ray 
flare at 54880~MJD with a doubling timescale of as short as 1 day. 
This sharp $\gamma$-ray flare coincides with a significant drop 
of the level of optical polarization (polarization degree: PD), 
from $\sim 30$\% down to a few per cent, lasting for $\Delta t \sim 20$ days. 
Subsequently, both $\gamma$-ray and optical fluxes gradually decrease together 
and reach the quiescent level, followed by a temporary recovery of the high degree of polarization.
This event is associated with a dramatic change of the electric vector 
position angle (EVPA) of the polarization, in contrast to being relatively constant 
before the event at $\sim$ 50\degree ~(parallel to the jet 
direction observed by Very Long Baseline Interferometry observations
in radio bands; see ref.~\citen{Jor05} for example).  Because the EVPA has 
$\pm180^{\circ}\times n$ (where $n=1,2...$) ambiguity, we selected values on the assumption of a smooth change 
of the EVPA, such that it would follow the overall trend.  
The polarization 
angle increases slightly at 54880~MJD -- coincident with the 
$\gamma$-ray flare -- then decreases by $208^{\circ}$ with a rate of 
$\sim 12^{\circ}$ per day, and returns to a level nearly exactly $180^{\circ}$ 
from the original level, resembling closely the behavior of optical 
polarization measured in BL~Lacertae~\cite{Mar08}, but at a rate
four times slower. This clearly indicates that the sharp $\gamma$-ray 
flare is unambiguously correlated with the dramatic change of optical 
polarization due to a single, coherent event, rather than 
a superposition of multiple but causally unrelated, shorter duration events.  

\bigskip

Concurrent X-ray observations indicate a relatively steady X-ray flux during the 
high $\gamma$-ray flux state 
(although with modest amplitude variations roughly mirroring 
the $\gamma$-ray time series;  A. Marscher, priv. comm.), 
but reveal a significant, symmetrical flare about 60 days 
after the second $\gamma$-ray peak --- 
at 54950~MJD --- with duration of $\sim 20$ days, similar to the duration 
of the $\gamma$-ray flare. 
It suggests the X-ray photons are produced at a distance 
from the black hole comparable to the distance of the optical/$\gamma$-ray photons.
Importantly, this X-ray flare is accompanied only by a modest increase 
of optical activity and not by a prominent optical or $\gamma$-ray flare.  
The X-ray spectrum during the isolated flare remains much harder than 
the optical spectrum (see Figure~2), 
and therefore cannot be attributed to a temporary extension of the high-energy tail of the 
synchrotron emission, but instead, may be generated by inverse-Compton 
scattering of low-energy electrons.  However, the similarity of profiles 
of the $\gamma$-ray and X-ray flares argues against the latter being just
a delayed version of the former due to, e.g., particle cooling.
Therefore, the X-ray flare must be produced independently by another mechanism 
involving primarily lower energy electrons. 

\bigskip

During the entire multiwavelength campaign reported here, 
the radio and millimeter fluxes are less variable than fluxes in other bands. 
In particular, they stay nearly constant in the periods of 
the two prominent $\gamma$-ray flares and the isolated X-ray flare, 
and no associated or delayed radio flare was observed.
This suggests that the blazar activity in 3C~279 takes place 
where the synchrotron radiation at these wavelengths is not yet 
fully optically thin, constraining the transverse size $R_{\rm blazar}$
of the blazar emission zone \cite{Sik94}

$$ R_{\rm blazar} < 5 \times 10^{16} \, (\nu F_{\nu}/2 \times 10^{-11}\,{\rm erg\,cm^{-2}\,s^{-1}})^{1/2}
\, (B'/0.3\,{\rm G})^{1/4} \, (\nu /10^{11.5}\,{\rm Hz})^{-7/4} \, (\Gamma_{\rm jet}/15)^{-1/4} 
\, {\rm cm} \, $$
(where $\nu F_{\nu}$ is the energy flux measured in the millimeter band [$\sim10^{11.5}$\,Hz]), 
which is consistent with 
the limit provided by shortest doubling timescales of the $\gamma$-ray flux variations.  

\bigskip

The gradual rotation of the polarization angle is unlikely to 
originate in a straight, uniform axially symmetric, matter-dominated jet 
because any compression of the jet plasma by, for example, a perpendicular shock moving along 
the jet and viewed at a small but constant angle to the jet axis 
would change the degree of polarization, but would not result 
in a gradual change of EVPA.  Instead, it can reflect a non-axisymmetric 
magnetic field distribution (as in, for example, ref.~\citen{Konig85}), a swing of the 
jet across our line of sight 
(which in turn does not require any source/pattern propagation), 
or a curved trajectory of the dissipation/emission 
pattern.  The last possibility may be due to propagation of 
an emission knot following a helical path in a magnetically dominated 
jet as was recently investigated in the context of the optical polarization 
event seen in BL Lacertae~\cite{Mar08} or may involve the 
``global'' bending of a jet.  
The magnetic field in the emission region is anisotropic 
(presumably concentrated in the plane of a shock or disturbance propagating 
along the jet), so the degree and angle of observed polarization then depends on 
the instantaneous angle $\theta$ of the direction of motion of the radiating material 
to the line of sight.  The maximum rotation rate of the 
polarization angle would correspond to $\theta = \theta_{\rm min}$ and
polarization degree would be highest for $\theta \sim 1/\Gamma_{\rm jet}$.
The ``bent jet'' scenario can explain 
the observed polarization event (the change of the angle as well as the 
magnitude of polarization) provided the jet curvature is confined to 
the plane inclined to the line of sight at an angle 
$\theta_{\rm min} < 1/\Gamma_{\rm jet}$ and configured in such a way that 
the jet trajectory projected on the sky turns by almost 180$^{\rm o}$.  
Similar geometry - albeit on larger scales - has 
been observed in another blazar, PKS 1510-089 (ref.~\citen{Hom02}).  
Nonetheless, in both scenarios, the coherent polarization event 
is produced by a density pattern co-moving along the jet, and therefore, 
it is possible to estimate the distance traveled by
the emitting material during the flare $\Delta r_{\rm event}$;  this in turn
allows us to constrain the distance $r_{\rm event}$ of the dissipation region (where 
flaring occurs) from the black hole, because $r_{\rm event} \geq \Delta r_{\rm event}$.  With this, 
$ r_{\rm event} \ge \Delta r_{\rm event} \sim 10^{19}\,(\Delta t_{\rm event}/20 
\,{\rm days})\,(\Gamma_{\rm jet}/15)^2 \, {\rm cm}\, , $
which is $\sim$ 5 orders of magnitude 
larger than the gravitational radius of the black hole in 3C~279.  

\bigskip

The constraints on the distance of the dissipation region
can be relaxed under 
``flow-through'' scenarios, in where the emission patterns may move much more slowly 
than the bulk speed of the jet or not propagate at all:  one such example is 
the model involving swings (``wobbling'') of the jet associated with jet instabilities 
such that its boundary moves relative to our line of sight.  In this case, the timescale for 
the observed variation is the timescale for the jet motion.  Consequently, 
the emission region easily can be much closer (by a factor $\Gamma_{\rm jet}^{2}$) 
to the black hole than in the ``helical'' or ``bent jet'' scenarios, 
because the natural radial scale for $\Delta t_{\rm event} \sim 20$ 
days is $r_{\rm event} \sim c \Delta t_{\rm event} \sim 500 - 1000$ 
gravitational radii (see, e.g., ref.~\citen{lyut03}).  Under this scenario, 
the angle the jet makes with the line of sight must change 
by at least $\sim \Gamma_{\rm jet}^{-1}$ in 
order to explain the large swing of polarization.  Here, the jet motion can be 
imposed at its base, be caused by deflection due to external medium, or be 
a consequence of dynamical instability. 

\bigskip

This leaves us with three viable possibilities.  
Both the scenario involving a knot propagating along the helical magnetic field lines 
and the ``flow-through'' scenario above imply that the rotation of the polarization angle 
should be preferentially following the same direction, because in those two 
models the twist presumably originates in the inner accretion disk. In our case, we observe the 
rotation of the polarization angle to be opposite in direction to that measured 
previously\cite{Lar08}, leaving us with the ``bent jet'' model 
combined with a small swing of the jet as the most compelling scenario.

\bigskip

The dominant source of ``seed'' photons for inverse-Compton scattering depends
on the distance of the dissipation event from the central black hole~\cite{Sik09}.  
At the parsec distances predicted by the ``helical'' or ``bent jet'' scenarios 
involving the radiating material 
co-moving with the jet, the ``seed'' radiation fields are dominated by infrared radiation 
emitted by a warm dust located in the circumnuclear molecular torus and by 
synchrotron radiation produced within the jet.  At sub-parsec distances 
implied by the ``flow-through'' scenarios, this photon field can be the 
broad emission line region~\cite{Sik94} (clearly detected in this 
object~\cite{Netz94}, as expected in a quasar possessing a luminous 
accretion disk~\cite{Pian99}), as well 
as the direct radiation of such a disk~\cite{Derm92} or its corona~\cite{BlandLev95}.   
In any case, the $\sim 20$ GeV electrons and positrons producing the 
highest-energy $\gamma$ rays and the polarized optical radiation lose their 
energy on timescales shorter than the light travel time from the black hole, 
and so must be accelerated locally.  

\bigskip

In summary, the close association of the energetically dominant 
$\gamma$-ray flare with the smooth, continuous change of the optical polarization angle 
suggests co-spatiality of the optical and $\gamma$--ray emission and provides evidence 
for the presence of highly ordered magnetic fields in the regions of the $\gamma$--ray 
production.  Provided the emission pattern is co-moving with the jet, we can 
measure the distance of the coherent event to be of the order 
of $10^{5}$ gravitational radii away from the black hole.  While the available data
cannot exclude the theoretically less explored ``flow-through'' 
scenarios - where the dissipation events may take place at much smaller distances, 
down to $\sim 10^{3}$ gravitational radii - the opposite direction of rotation of the 
optical polarization angle than previously measured appears to support the 
jet bending at larger distances as the best explanation of the available data.  
Furthermore, the detection of the isolated X-ray flare challenges 
the simple, one-zone emission models, rendering them too simple.  Regardless, 
the \Fermi\ satellite has been in operation for only just over a year, and the outlook for 
a more comprehensive picture of these enigmatic objects, primarily 
via multi-band campaigns including well-sampled optical polarimetry, is excellent.


\noindent {\bf Acknowledgments:} The {\it Fermi}-LAT Collaboration acknowledges support from a number of agencies and institutes for both development and the operation of the LAT as well as scientific data analysis. These include NASA and DOE in the United States, CEA/Irfu and IN2P3/CNRS in France, ASI and INFN in Italy, MEXT, KEK, and JAXA in Japan, and the K.~A.~Wallenberg Foundation, the Swedish Research Council and the National Space Board in Sweden. Additional support from INAF in Italy for science analysis during the operations phase is also gratefully acknowledged. 
The GASP-WEBT observatories participating in this work are Abastumani, Calar Alto, Campo Imperatore, Crimean, Kitt Peak (MDM), L'Ampolla, Lowell (Perkins-PRISM), Lulin, Roque de los Muchachos (KVA and Liverpool), San Pedro M\'artir, St.\,Petersburg for the optical-NIR bands, and Mauna Kea (SMA), Medicina, Metsahovi, Noto and UMRAO for the mm-radio band, and are supported in part by the Georgian National Science Foundation, the Spanish ``Ministerio de Ciencia e Innovaci\'on", the NSF and NASA and the Smithsonian Institution in the United States,  the UK Science and Technology Facilities Council,  the Academia Sinica in Taiwan, the Russian RFBR and the Academy of Finland. M.~Hayashida is supported by the JSPS for the Postdoctoral Fellowship for Research Abroad.
\\

\noindent {\bf Author Contributions:} 
All authors contributed to the work presented in this paper.
The Kanata observations and data analysis
were led by M.~Uemura and R.~Itoh. M.~Villata organized
the optical-radio observations by GASP-WEBT as the president
of the collaboration.
\\

\noindent {\bf Author Information:} Correspondence and requests for materials should be addressed to 
M.~Hayashida (mahaya@slac.stanford.edu) and G.~Madejski (madejski@slac.stanford.edu)

\medskip

\newpage

\noindent \underline{ \bf Figure Legends}

\smallskip 

\noindent 
{\bf Figure 1: History of flux in various bands, $\gamma$-ray photon index, and 
optical polarization of 3C~279.} 
Light curves at the indicated wave bands covering 1 year since the Modified Julian Day (MJD) of 54650 (corresponding to July 3rd 2008).
The two dashed vertical lines indicate 54880 and 54900~MJD.  
Error bars at each point represent a $\pm1$ s.d.~statistical uncertainty. 
{\bf a-b,} Gamma-ray flux $F_{\gamma}$ and photon 
index {$\mathit \Gamma$} above 200 MeV averaged over 3-day intervals as measured 
by \Fermi-LAT based on photons that passed the ``diffuse'' event selection.
The source fluxes are calculated using "P6\_V3\_DIFFUSE" for the instrumental 
response function and a simple power-law spectral model (${\rm d}F/{\rm d}E \propto E^{-\mathit \Gamma}$).  
 The detailed data analysis procedures are analogous 
to those in ref.~\citen{LAT_3C454.3}.
{\bf c,} X-ray integrated flux $F_{\rm X}$ between 2 and 10 keV, calculated by 
fitting the data with the simple power-law model taking into account a Galactic 
absorption. Light-green points are from the observations with the 
Proportional Counter Array (PCA) on-board the Rossi X-ray Timing Explorer (\RXTE) and 
dark green points are measurements by \Swift-XRT.
{\bf d,} Optical and ultra-violet (UV) fluxes
in several bands. $R$-band data were taken by ground-based 
telescopes from the GASP-WEBT collaboration\cite{Vil09}. 
$V$-band data were taken by a 
ground-based telescope (Kanata-TRISPEC~\cite{Kanata}) and \Swift-UVOT.  
Data in all other bands were acquired by \Swift-UVOT. 
{\bf e-f,} Polarization degree 
(PD) and electric vector position angle (EVPA) of the optical polarization 
measured by the Kanata-TRISPEC in the $V$-band 
(dark blue) and by the KVA telescope without any filters (light blue). Note 
that EVPA has $\pm180^{\circ}\times n$ (where $n=1,2...$) ambiguity. 
The horizontal dashed lines in {\bf f} refer to EVPA of $50$\degree and $-130$\degree.
{\bf g-h,} Near-infrared (NIR) and radio fluxes measured by ground-based telescopes 
(Kanata-TRISPEC [$J$, $K_s$], OVRO [15~GHz] and GASP-WEBT 
[$J, H, K$ and  several millimeter and radio bands]). All UV, 
optical and NIR data are corrected for the Galactic absorption.

\smallskip  

\noindent  {\bf Figure 2: Energy spectrum from radio to $\gamma$-ray band of 3C~279 
at two different epochs.}  The red points were taken between 54880 and 54885~MJD, 
corresponding to the first five days of the sharp $\gamma$-ray 
flare accompanying the dramatic polarization 
change event [epoch-1].  The blue points were taken between 54950 and 54960~MJD, 
around the peak of the isolated X-ray flare [epoch-2].
The $\gamma$-ray spectra were measured by \Fermi-LAT. 
In the X-ray band, the flux points are obtained by the \RXTE-PCA in 
the epoch-1 (red) and by \Swift-XRT in the epoch-2 (blue). 
The fluxes in the UV range were measured by \Swift-UVOT.
Observations in the optical-to-radio bands were performed 
by ground-based telescopes as given in Figure 1 (with additional radio 
coverage provided by the Effelsberg radio telescope\cite{Lars}).   
Each data point represents an average source flux and the 
error bar represents $\pm1$ s.d.~of the flux 
during each epoch. Each data point is already corrected for Galactic absorption.
Note that the total energy associated with the X-ray flare is 
relatively modest, about 30 times less than the energy associated with 
the $\gamma$-ray flare accompanying the dramatic polarization 
change, and the $\gamma$-ray emission is still dominant, 
having five times the X-ray energy flux even during the X-ray flare event.

\newpage 

\begin{figure}[htbp]
\begin{center}
\includegraphics[width=12cm]{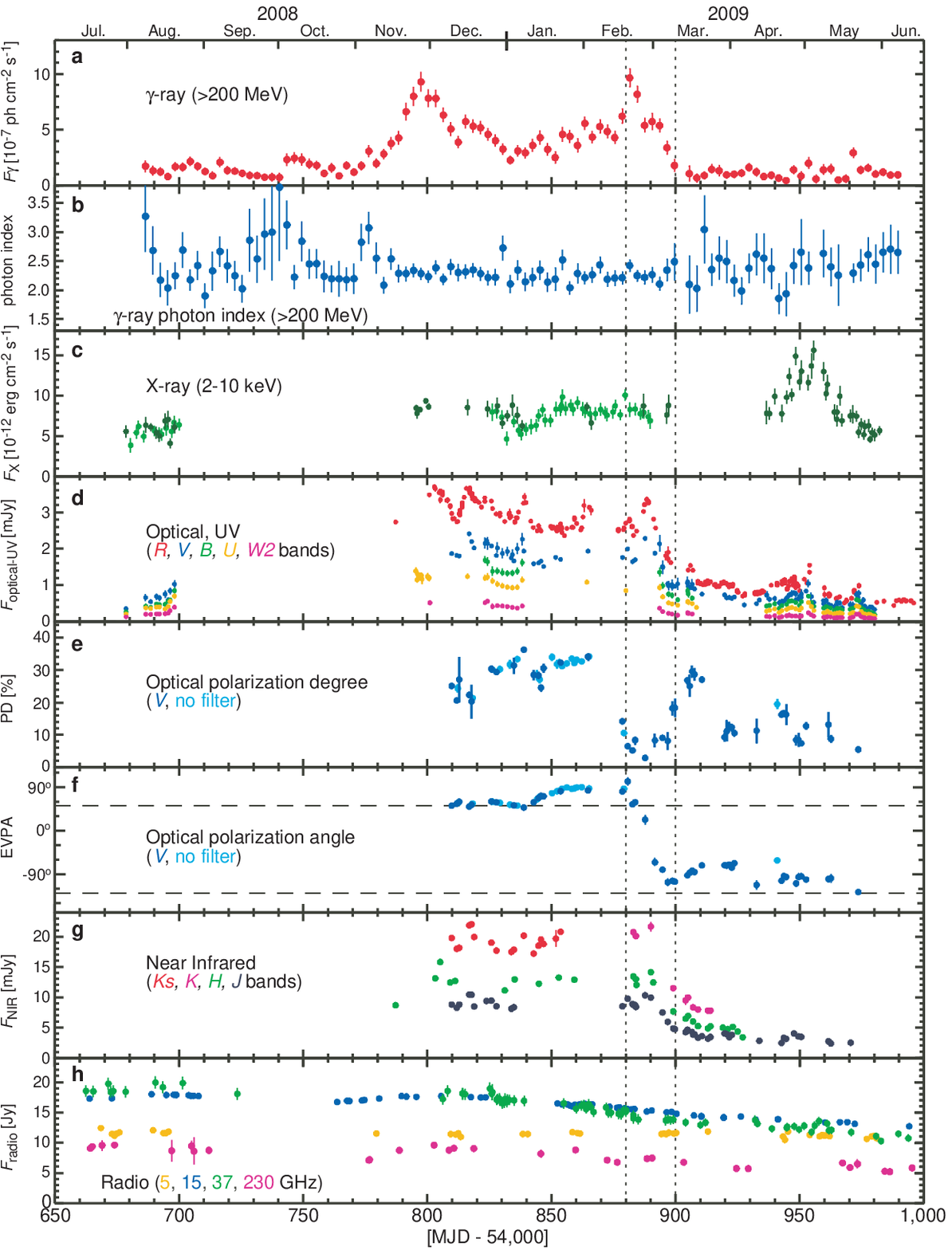}
\caption{}
\label{fig:LC}
\end{center}
\end{figure}

\newpage 

\begin{figure}[htbp]
\begin{center}
\includegraphics[width=8.9cm]{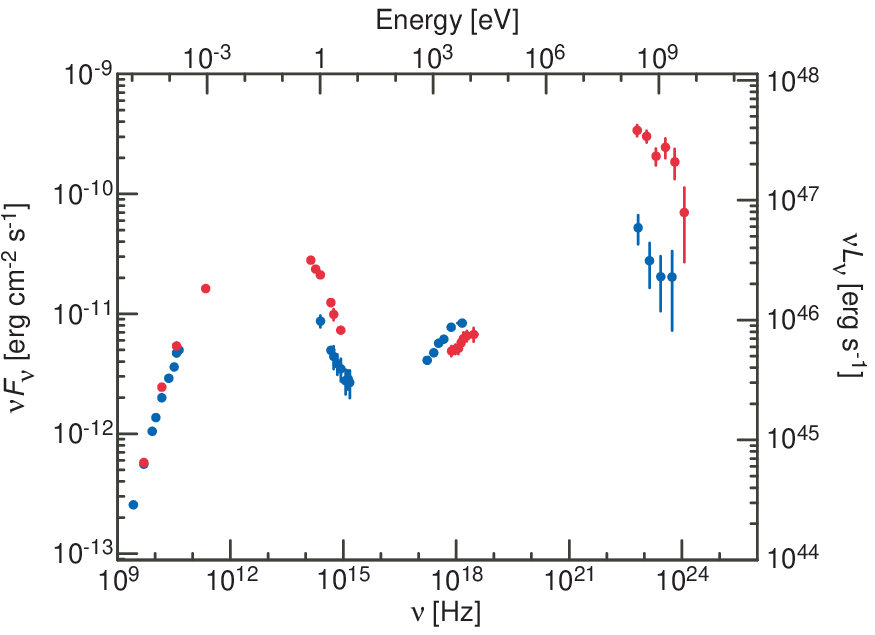}
\caption{}
\label{fig:SED}
\end{center}
\end{figure}
\newpage 
\end{document}